\documentclass[journal]{IEEEtran}
\usepackage{fancyhdr}
\usepackage{amsmath,amssymb}
\usepackage{amsfonts}
\usepackage[dvips]{graphicx}
\usepackage{dsfont}
\usepackage{balance}
\usepackage{algpseudocode}
\usepackage{algorithm}

\usepackage[hidelinks]{hyperref}    

\usepackage{color}
\usepackage{pgfplots}           
\pgfplotsset{compat=1.16}
\usepackage{enumerate}
\usepackage{graphics}
\usepackage{subfigure}
\usepackage{cite}
\usepackage{mathrsfs}
\usepackage{stfloats}

\usepackage{tikz}
\usepackage{mathdots}
\usepackage{yhmath}
\usepackage{cancel}
\usepackage{color}
\usepackage{siunitx}
\usepackage{array}
\usepackage{multirow}
\usepackage{textcomp}
\usepackage{gensymb}
\usepackage{tabularx}
\usepackage{booktabs}
\usepackage{graphicx}
\usepackage{xcolor}

\usepackage{enumitem}

\newtheorem{proposition}{Proposition}

\makeatletter
\newcommand*{\rom}[1]{\expandafter\@slowromancap\romannumeral #1@}
\makeatother

\begin{document}

\title{\vspace{-1cm}  Performance Analysis of Cascaded Reconfigurable Intelligent Surface Networks}
\author{
Dimitrios Tyrovolas,~\IEEEmembership{Student Member,~IEEE,}
Sotiris A. Tegos,~\IEEEmembership{Student Member,~IEEE,}\\
Emmanouela C. Dimitriadou-Panidou,~\IEEEmembership{Student Member,~IEEE,}\\
Panagiotis D. Diamantoulakis,~\IEEEmembership{Senior Member,~IEEE,}
Christos K. Liaskos,~\IEEEmembership{Member,~IEEE,}\\
and George K. Karagiannidis,~\IEEEmembership{Fellow,~IEEE}
\thanks{ D. Tyrovolas, S. A. Tegos, E. C. Dimitriadou-Panidou, P. D. Diamantoulakis and G. K. Karagiannidis are with the Wireless Communications and Information Processing (WCIP) Group, Electrical \& Computer Engineering Dept., Aristotle University of Thessaloniki, 54 124, Thessaloniki, Greece (e-mails: \{tyrovolas,tegosoti,emmadimi,padiaman,geokarag\}@auth.gr).
C. K. Liaskos is with the Computer Science Engineering Department, University of Ioannina,
45110 Ioannina, Greece (e-mail: cliaskos@cse.uoi.gr).}
}
\maketitle
%
\vspace{-1.5cm}
\begin{abstract}
Reconfigurable intelligent surfaces (RIS) have been presented as a solution to realize the concept of smart radio environments, wherein uninterrupted coverage and extremely high quality of service can be ensured. In this paper, assuming that multiple RIS are deployed in the propagation environment, the performance of a cascaded RIS network affected by imperfect phase estimation is evaluated. Specifically, we derive closed-form expressions for the outage probability, the ergodic capacity and the average symbol error probability that can be utilized to evaluate the coverage of the proposed network, as well as the average capacity and the data transmission accuracy. Finally, we validate the derived expressions through simulations and show that by choosing the number of the participating RIS correctly, a cascaded RIS network can outperform a single RIS-aided system and extend the network’s coverage efficiently.
\end{abstract}


\IEEEpeerreviewmaketitle
\vspace{-0.3cm}
\section{Introduction}
Future wireless networks are expected to play a pivotal role in society as they will offer access to intelligent applications such as autonomous driving, virtual and augmented reality etc. \cite{6G}. In order to offer ubiquitous services, though, wireless connectivity should be provided for everyone and everywhere \cite{SRE}.  In this context, reconfigurable intelligent surfaces (RISs) have been presented as a solution to realize the concept of smart radio environments (SREs) in which uninterrupted coverage and extremely high quality of service (QoS) can be ensured \cite{alouwu,path}. In more detail, RISs are thin programmable surfaces consisting of reflecting elements, whereby tuning their phase profile, the desired reflection coefficient is constructed leading to the realization of different electromagnetic functions (i.e., beamforming, diffusion etc.). Therefore, to enhance the wireless network's coverage, a RIS can be appropriately configured in order to act as a passive beamformer and steer its impinging signal towards the desired direction \cite{liaskos,tegos}. Thus, by deploying multiple RISs across the propagation environment, it can be converted from an uncontrollable entity to an optimizable parameter, able to provide wireless connectivity seamlessly \cite{tutorial}.

Assuming that multiple RISs are deployed within the wireless propagation environment, multi-RIS wireless networks' performance should be examined to determine their capabilities. \color{black} In a distributed multi-RIS network, the RISs are densely deployed across the propagation environment in order to enhance the communication's QoS by serving each ground node (GN) from diverse communication links \cite{rev}. \color{black} Specifically, in \cite{Haas}, the performance of a distributed multi-RIS-aided wireless network with perfect phase estimation was investigated, where each RIS is used to steer its impinging signal towards a GN. \color{black} However, if a GN can form a link with only one RIS that cannot form a link with the BS due to the propagation environment, an alternative RIS network is needed that can bypass any existing obstacles and extend the network's coverage efficiently. Besides, distributed RIS networks require dense RIS deployment within the environment to serve every GN with more than one RIS, thus, becoming less cost-efficient.

Cascaded RIS networks are intended to efficiently bypass obstacles within the propagation environment and serve the GN through a unique cascaded communication link where each RIS is selected to either steer the signal towards another RIS or towards the GN \cite{zhangcoop}. Specifically, a cascaded RIS network can be used to extend the network's coverage by creating new links between the GN and the BS through a RIS-enabled route that can be determined via routing algorithms \cite{route}. As shown in \cite{route}, the squared channel gain in cascaded RIS networks is proportional to the squared product of the RISs' reflecting elements, proving that cascaded RIS networks achieve more significant gain than single RIS-aided systems.  \color{black} However, most of the existing works in cascaded RIS networks assume that every communication link between each node in the RIS-route is not affected by small-scale fading and that perfect phase estimation exists \cite{zhangcoop}, \cite{route}. This assumption, though, can be proven impractical due to the possible existence of scatterers between the BS and the first RIS of the route and the GN's non-static nature. Therefore, due to the existence of small-scale fading and the complexity of perfect phase estimation in a cascaded-RIS network,  imperfect phase estimation needs to be taken into consideration \cite{uaval}. To the best of the authors' knowledge, the performance of a cascaded RIS system affected by small-scale fading with a various number of RISs and imperfect phase estimation has not yet been examined.

\color{black}This paper evaluates the performance of a cascaded RIS wireless network affected by imperfect phase estimation in a downlink scenario between a single-antenna BS and a single-antenna GN which has perfect knowledge of the channel. Specifically, we assume that the estimation procedure consists of phase estimation in order to configure the RISs and after the RISs configuration, the GN estimates the end-to-end channel perfectly. It should be noted that the assumption of perfect knowledge of the end-to-end channel is practical because only its amplitude and phase need to be estimated. After the estimation procedure, the BS transmits towards a specified RIS and then the GN is served via a unique cascaded RIS-route. \color{black} To characterize the performance of a cascaded RIS system, we derive closed-form expressions for the outage probability (OP), the ergodic capacity (EC), and the average symbol error probability (ASEP) for different modulation schemes. Finally, we validate the derived expressions through simulations where it can be observed that if the number of participating RIS is chosen appropriately, a cascaded RIS network can outperform a single-RIS system even if the phase estimation's accuracy is worse than that of a single-RIS system. It should be mentioned that the derived results can be utilized to develop routing algorithms which choose the best RIS-route to maximize the communication performance of a specific GN.
\begin{figure}
\centering
\includegraphics[width=1.00\linewidth]{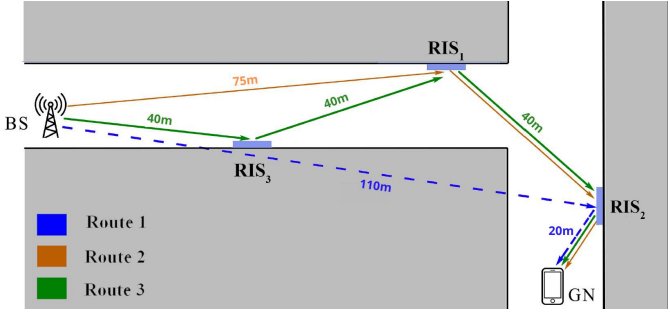}
  \caption{A cascaded RIS communication network.}
  \label{fig:cascaded}
\end{figure}

\section{System Model}\label{sysmodl}
We consider a cascaded RIS network in an urban environment consisting of a BS and a GN, both equipped with a single antenna.  \color{black} Specifically, $L \geq 2$ RISs are deployed across the propagation environment, and a unique RIS-route is selected to serve a GN which cannot form a direct communication link with the BS due to the propagation environment's harshness. \color{black}  Also, according to the main principle of cascaded RIS networks, it is assumed that the BS's antenna is sufficiently directive, such as solely the first RIS of the route belongs in its field-of-view. Indicatively, in Fig. 1, a scenario with three RISs is illustrated, in which solely Route 2 and Route 3 achieve line-of-sight links between the involved nodes. It is noted that this indicative example is also used for the evaluation of cascaded RIS networks in Section \ref{secnum}. \color{black} It should be highlighted that even if a small amount of power impinges upon a RIS that does not participate in the selected route or is not configured properly, the impinging power will be randomly reflected across the space due to the random configuration of the RISs, thus, the GN can be served only via the selected RIS-route \cite{tegos}. \color{black}

For the proposed communication scenario, the received signal at the GN can be modeled as
\begin{equation}\label{signal}
Y_c =    \sqrt{l_c G P_t} H_c X + W ,
\end{equation}
where $X$ is the transmitted signal for which it is assumed that $\mathbb{E}[|X|^2]=1$ with $\mathbb{E}[\cdot]$ denoting expectation. Furthermore, $W$ denotes the additive white Gaussian noise which is a complex Gaussian random variable (RV), i.e., $W \sim \mathcal{CN}\left(0,{\sigma}^2 \right)$ where ${\sigma}^2$ is the noise power. Also, $H_c$,  $P_t$ and $G = G_t G_r$  denote the equivalent channel, the transmit power and the product of the BS and the GN antenna gains, respectively. Moreover, $l_c$ denotes the total route's path loss which can be modeled as
\begin{equation}\label{pathloss}
l_c  = \prod_{i=1}^{L+1} C_0 \left( \frac{d_i}{{d_0}} \right)^{-n_i},
\end{equation}
where $d_{i}$ is the distance between the $i$-th and the $\left(i+1\right)$-th node of the route, i.e., the $i$-th link of the route, $C_0$ is the path loss at the reference distance $d_0$ and $n_i$ is the path loss exponent of the $i$-th link. 

It should be highlighted that to improve the quality of the end-to-end communication link, the RISs are correctly placed on establishing LoS channels with each other. Therefore, it can be assumed that the communication links between neighboring RIS are not affected by small-scale fading. Finally, the channels between the BS and the first RIS of the RIS-route and between the last RIS and the GN undergo Nakagami-$m$ fading and due to imperfect phase estimation is assumed that the first and the last RIS do not correct the phase of the impinging signal ideally.




\section{Performance Metrics}
In this section, we present the OP, the EC as well as the ASEP for a wireless network with $L$ cooperating RISs, which is affected by imperfect phase estimation. In order to calculate the aforementioned metric, the equivalent channel $H_c$ needs to be characterized
\begin{proposition}
 The channel ${H_c}$ for a cascaded RIS wireless network where its first and last RIS are imperfectly configured due to imperfect phase estimation, can be expressed as

\begin{equation}\label{casc}
{H_c} = \prod_{i=2}^{L-1} N_i \sum_{i=1}^{N_1} h_{1i}  e^{-j \phi_{1i}} \sum_{i=1}^{N_L} h_{Li}  e^{-j \phi_{Li}} ,
\end{equation}
where $N_l$ with $l = \{1, \ldots, L\}$ is the reflecting elements number of the $l$-th RIS, $h_{ki}$ with $k = \{1, L\}$ denotes the channel coefficient following the Nakagami-$m$ distribution with shape parameter $m_{k}$ and spread parameter $\Omega_{k}$, \color{black} and $\phi_{ki}$ denotes the phase error term of the $i$-th element of the $k$-th RIS which follows the von Mises distribution with concentration parameter $\kappa_k$, where $\kappa_k$ is inversely proportional to the mean-squared error (MSE) of the phase estimation and indicates the phase estimation accuracy. It should be highlighted that the Nakagami-$m$ distribution is a general case able to describe accurately the cases where LoS component is included or not, allowing us to extract useful insights.\color{black}
\end{proposition}



\begin{IEEEproof}
The equivalent channel $H_c$ for a cascaded RIS system can be expressed as follows
\begin{equation}
{H_c} = \boldsymbol{h_{L+1}} \mathrm{diag}(\boldsymbol{\Phi_L}) \ldots \boldsymbol{h_2} \mathrm{diag}(\boldsymbol{\Phi_1}) \boldsymbol{h_1},
\end{equation}
where $\boldsymbol{h_l} \in \mathbb{C}^{ N_{l+1} \times N_l}$ are communication channels undergoing Nakagami-$m$ fading with shape parameter $m_{l}$ and spread parameter $\Omega_{l}$ and $\boldsymbol{\Phi_l} = [ e^{j \theta_1},\ldots,  e^{j \theta_{N_{l}}}]^{\boldsymbol{T}} \in \mathbb{C}^{N_{l}}$ denotes the phase shift vector of the $l$-th RIS with $\theta_l \in [0, 2\pi]$. \color{black} Considering that the communication channels between the RISs are not affected from small-scale fading (i.e. $h=1$) and that their phase is a deterministic variable equal to $\frac{2 \pi r}{\lambda}$, where $r$ is the euclidean distance between the RISs and $\lambda$ is the wavelength, the equivalent channel $H_c$ can be rewritten as \color{black}
 \begin{equation}
H_c = \sum_{i_1=1}^{N_1} h_{1i_1} e^{-j \omega_{1i_1}} \sum_{i_2=1}^{N_2} \ldots \sum_{i_{L-1}=1}^{N_{L-1}} \sum_{i_L=1}^{N_L} h_{Li_L} e^{-j \omega_{Li_L}},
\end{equation}
where $ \omega_{ki_k} =  \arg(h_{ki_k}) +  \theta_{i_k}$ with $\mathrm{arg}(\cdot)$ denoting the argument of a complex number. By taking into account that the phases of the channels between the BS and the $i$-th element of the first RIS of the route and between the $i$-th reflecting element of the last RIS of the route and the GN are not estimated perfectly, the term $\omega_{ki_k}$ as mentioned in \cite{justin} can be expressed as a von Mises distributed RV, leading to (\ref{casc}) which concludes the proof. \color{black} It should be mentioned that every RIS can steer perfectly its impinging signal towards its neighboring RISs, due to the fact that the phases of the channels between the RISs are perfectly known as they depend only on the euclidean distance between the neighbouring RISs which is precisely known. \color{black}
\end{IEEEproof}

 Assuming that the number of the reflecting elements of each RIS is large, according to the central limit theorem each summation term in (\ref{casc}) can be written as a random vector with its real and imaginary parts being normally distributed RVs where its amplitude can be tightly approximated with an RV $Z_k$, following the Nakagami-$m$ distribution with spread parameter $\bar{\Omega}_k$ and shape parameter $\bar{m}_k$ which are given, respectively, by
\begin{equation}
	\bar{\Omega}_k = {t_1}^2 \left( \frac{\Gamma(m_{k}+0.5)}{\Gamma(m_{k})}  \sqrt{\frac{\Omega_{k}}{m_{k}}} \right)^2, 
\end{equation}
and
\begin{equation}
	\bar{m}_k = \frac{ N_k \bar{\Omega}_k}{2 + 2 t_2 - 4\bar{\Omega}_k},
\end{equation}
where $t_n = \mathbb{E} [ e^{jn \phi_k}]$ is the $n$-th trigonometric moment of $\phi_k$, which can be expressed as
\begin{equation}\label{pros}
t_n = \dfrac{I_n \left( \kappa_k \right)}{I_0 \left(  \kappa_k \right)}
\end{equation}
with $I_p(\cdot)$ being the modified Bessel function of the first kind and order $p$ \cite{justin}. Thus, the amplitude of $H_c$ can be approximated as 
\begin{equation}
\lvert H_c \rvert \approx \prod_{i=1}^{L} N_i Z_1 Z_L  ,
\end{equation}
and the instantaneous received SNR at the GN of the proposed multi-RIS system can be expressed as

\begin{equation}\label{av}
\gamma_r \approx  \frac{P_t}{\sigma^2} l_c G \left( \prod_{i=1}^{L} N_i Z_1 Z_L \right)^2.
\end{equation}


\subsection{Outage probability}
In order to evaluate the coverage of the proposed network, we calculate the OP which is defined as the probability that the instantaneous received SNR is below a specified threshold.
\begin{proposition}
The OP of the presented communication system can be expressed as
\begin{equation}\label{OP}
\mathcal{P}_o \approx \frac{1}{\Gamma(\bar{m}_1) \Gamma(\bar{m}_L)} G_{1,3}^{2,1} \left( w^2 \frac{\bar{m}_1\bar{m}_L}{\bar{\Omega}_1\bar{\Omega}_L} | 					
                                                     \begin{array}{c}
						1 \\
						\bar{m}_1, \bar{m}_L, 0 \\
					\end{array} \right),
\end{equation}
where $G_{m,n}^{p,q} (\cdot | \cdot)$ is the Meijer-G function, $\Gamma(\cdot)$ is the gamma function, $w = \sqrt{ \frac{\gamma_{\mathrm{thr}} {\sigma}^2}{P_t l_c G \prod_{i=1}^{L} {N_i}^2 }}$ and $\gamma_{\mathrm{thr}}$ is the outage threshold value for the received SNR.
\end{proposition}
\begin{IEEEproof}
The OP for the proposed multi-RIS system can be defined as
\begin{equation}
\begin{split}
\mathcal{P}_o &= \Pr \left(\gamma_r \leq \gamma_{\mathrm{thr}} \right) \approx \Pr \left( Z_1 Z_L \leq {w} \right).
\end{split}
\end{equation}
Considering that $Z_1$ and $Z_L$  are i.i.d Nakagami-$m$ distributed RVs, the OP can be calculated through the cumulative density function of double-Nakagami-$m$ distribution which is constructed as the product of two statistically independent but not necessarily identically distributed, Nakagami-$m$ RVs \cite{nakagami}. Thus, the OP can be expressed as in (\ref{OP}), which concludes the proof.
\end{IEEEproof}
\vspace{-0.5cm}
\subsection{Ergodic Capacity and Average Symbol Error Probability}
Next, we provide closed-form approximations for both EC and ASEP of the considered system in order to characterize the system's performance in terms of average channel capacity and accuracy of data transmission.
\begin{proposition}
The EC of the considered network can be approximated as
\begin{equation}\label{EC}
\begin{split}
C_c & \approx \frac{1}{ \mathrm{ln}2 \Gamma(\bar{m}_1) \Gamma(\bar{m}_L) } \\
& \times  G_{4,2}^{1,4} \left( \frac{P_t G l_c \prod_{i=1}^{L} {N_i}^2 }{ {\sigma}^{2} \frac{\bar{m}_1\bar{m}_L}{\bar{\Omega}_1 \bar{\Omega}_L} } | 					
                                                     \begin{array}{c}
						1,1,1-\bar{m}_1,1-\bar{m}_L \\
						1,0 \\
					\end{array} \right),
\end{split}
\end{equation}
where $\mathrm{ln}(\cdot)$ is the natural logarithm.
\end{proposition}
\begin{IEEEproof}
The EC of the proposed system is defined as
\begin{equation}\label{start}
C_c = \int_{0}^{\infty} \mathrm{log}_2 \left(1 + x^2\right) f_{\lvert {H_c} \rvert}(x) dx,
\end{equation}
where $f_{\lvert {H_c} \rvert}(\cdot)$ can be approximated by the probability density function (PDF) of the double-Nakagami-$m$ distribution \cite{nakagami}. \color{black} By converting $\mathrm{log}_2(\cdot)$ into a natural logarithm form and transforming the functions within the integral into Meijer-G functions, (\ref{EC}) can be derived through the results provided in \cite{Adamchik}, which concludes the proof. \color{black}
\end{IEEEproof}
\begin{table}
	\centering
	\caption{Values of $\alpha$ and $\beta$  for different modulation schemes.}
	\begin{tabular}{|c|c|c|}
		\hline
		Modulation & $\alpha$ & $\beta$ \\
		\hline 	\hline
		BPSK & $1$ & ${2}$ \\
		\hline
		QPSK & $2$ & $1$ \\
		\hline
		M-PSK & $2$ & $2 \mathrm{sin}^{2}(\frac{\pi}{M})$ \\
		\hline
		M-QAM & $4$ & $\frac{3}{M-1}$ \\
		\hline
	\end{tabular} 
	\label{Table1} 
\end{table}

In the following proposition, the ASEP is approximated.
\begin{proposition}
An ASEP approximation for a cascaded RIS system for many known modulation schemes (e.g. M-PSK, M-QAM) can be calculated as
\begin{equation}\label{asep}
\begin{split}
A_c & \approx \frac{\alpha}{2 \sqrt{\pi}  \Gamma(\bar{m}_1) \Gamma(\bar{m}_L)}  \\
& \times G_{3,2}^{2,2} \left( \frac{\beta P_t G l_c \prod_{i=1}^{L} {N_i}^2}{2 {\sigma}^{2} \frac{\bar{m}_1\bar{m}_L}{\bar{\Omega}_1 \bar{\Omega}_L}} | 					
                                                     \begin{array}{c}
						1-\bar{m}_1,1-\bar{m}_L,1 \\
						0,\frac{1}{2} \\
					\end{array} \right),
\end{split}
\end{equation}
where $\alpha$ and $\beta$ are modulation-dependent parameters which are presented in Table I \cite{goldsmith}.
\end{proposition}
\begin{IEEEproof}
The ASEP for many known modulation schemes according to \cite{goldsmith} can be expressed as
\begin{equation}
A_c = \int_{0}^{\infty} \alpha Q\Big(\sqrt{\beta}x\Big) f_{\lvert {H_c} \rvert}(x) dx,
\end{equation}
where $Q\left(x\right) = \frac{1}{2} \mathrm{erfc}\left(\frac{x}{\sqrt{2}}\right)$ and $\mathrm{erfc}(\cdot)$ is the complementary error function. \color{black} By  setting $x=\sqrt{z}$ and converting the functions within the integral into Meijer-G functions, by using \cite{Adamchik}, (\ref{asep}) is obtained, which concludes the proof. \color{black}
\end{IEEEproof}


\section{Numerical Results}\label{secnum}
\subsection{Simulation setup}
We examine the derived results in a downlink scenario affected by imperfect phase estimation, where a GN with an omnidirectional antenna, i.e., $G_r=1$, is served from a BS with a directional antenna with gain $G_t=10$, through a cascaded RIS-route, as illustrated in Fig. \ref{fig:cascaded}. Specifically, the communication channels between the BS and $\mathrm{RIS}_1$ and between $\mathrm{RIS}_2$ and the GN undergo Nakagami-$m$ fading and have shape parameters $m_1=m_L=3$, spread parameters equal to 1 and path loss exponent $n=2.4$. In contrast, the path loss exponent for the channels between each RIS is equal to $n=2$ due to the proper placement of the RISs. Moreover, both the concentration parameters of ${\phi_1}$ and $\phi_L$ are chosen to be equal, i.e., $\kappa_1=\kappa_L=\kappa$. Finally, unless it is stated otherwise, each RIS consists of 500 elements, the outage SNR value $\gamma_{\mathrm{thr}}$ is set at $10$ dB, $C_0$ is set at $-41$ dB for reference distance $d_0 = 1$ m and the noise power ${\sigma^2}$ is set at $-144$ dB.
\subsection{Performance Evaluation}

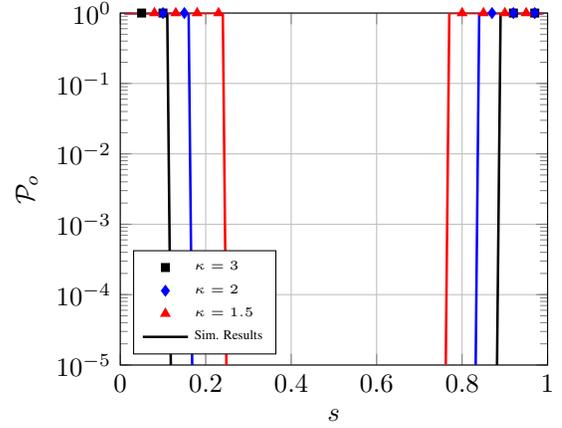
\begin{figure}
	\centering
	\begin{tikzpicture}
	\begin{semilogyaxis}[
	width=0.82\linewidth,
	xlabel = {$s$},
	ylabel = {$\mathcal{P}_{o}$},
	xmin = 0,xmax = 1,
	ymin = 0.00001,
	ymax = 1,
	xtick = {0,0.2,...,1},
	grid = major,
      legend style = {font = \tiny},
	legend cell align = {left},
	legend pos = south west
	]
	\addplot[
	black,
          only marks,
	mark=square*,
	mark repeat = 5,
	mark size = 1.5,
          mark phase = 5,
	]
	table {final/outr_3.dat};
	\addlegendentry{$\kappa=3$}	
	\addplot[
	blue,
	only marks,
	mark=diamond*,
	mark repeat = 5,
	mark size = 2,
          mark phase =10.5,
	]
	table {final/outr_2.dat};
	\addlegendentry{$\kappa=2$}	
	\addplot[
	red,
	only marks,
	mark=triangle*,
	mark repeat = 5,
	mark size = 2,
           mark phase =8,
	]
	table {final/outr_15.dat};
	\addlegendentry{$\kappa=1.5$}	
	\addplot[
	black,
          no marks,
	line width = 0.95pt,
	style = solid,
	]
	table {final/outr_3.dat};
	\addlegendentry{Sim. Results}	
	\addplot[
	blue,
	no marks,
	line width = 0.95pt,
	style = solid,
	]
	table {final/outr_2.dat};
	\addplot[
	red,
	no marks,
	line width = 0.95pt,
	style = solid,
	]
	table {final/outr_15.dat};
	\end{semilogyaxis}
	\end{tikzpicture}
	\caption{OP versus splitting factor $s$.}
	\label{fig:outage_el}
\end{figure}

In Fig. \ref{fig:outage_el}, it is illustrated how the number of the reflecting elements affects the OP in a double RIS network affected by imperfect phase estimation. In more detail, it is assumed that the GN is served via Route 2, as illustrated in Fig. 1, which consists of $\mathrm{RIS}_1$ and $\mathrm{RIS}_2$ with $N_1$ and $N_2$ reflecting elements, respectively, with $N_1+N_2=1000$, $N_1=\lfloor 1000 (1- s) \rfloor$ and $N_2=\lfloor 1000 s \rfloor$, where the splitting factor $s \in [0,1]$ and $\lfloor \cdot \rfloor$ is the floor operator. Finally, the transmit power $P_t$ is set equal to $20$ dBm, and the distances between the route's nodes are given as shown in Fig. \ref{fig:cascaded}. As it can be observed, as the value of $\kappa$ increases, the cooperation of RISs, even with a different number of reflecting elements, can enhance the network's coverage and reliability.
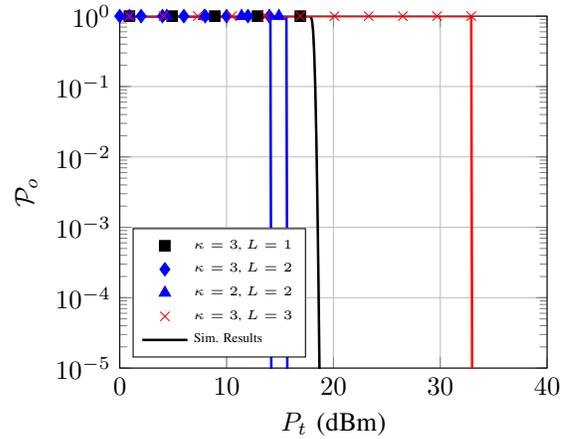
\begin{figure}
	\centering
	\begin{tikzpicture}
	\begin{semilogyaxis}[
	width=0.82\linewidth,
	xlabel = {$P_t$ (dBm)},
	ylabel = {$\mathcal{P}_{o}$},
	xmin = 0,xmax = 40,
	ymin = 0.00001,
	ymax = 1,
	xtick = {0,10,...,60},
	grid = major,
      legend style = {font = \tiny},
	legend cell align = {left},
	legend pos = south west
	]
	\addplot[
	black,
	only marks,
	mark=square*,
	mark repeat = 40,
	mark size = 2,
	mark phase=10,
	]
	table {final/out1_3.dat};
	\addlegendentry{$\kappa=3$, $L=1$}	
	\addplot[
	blue,
	only marks,
	mark=diamond*,
	mark repeat = 20,
	mark size = 2.5,
	]
	table {final/out2_3.dat};
	\addlegendentry{$\kappa=3$, $L=2$}	
	\addplot[
	blue,
	only marks,
	mark=triangle*,
	mark repeat = 35,
	mark phase= 10,
	mark size = 2.5,
	]
	table {final/out2_2.dat};
	\addlegendentry{$\kappa=2$, $L=2$}	
	\addplot[
	red,
	only marks,
	mark=x,
	mark repeat = 32,
	mark phase=10,
	mark size = 2.5,
	]
	table {final/out3_3.dat};
	\addlegendentry{$\kappa=3$, $L=3$}	
	\addplot[
	black,
	no marks,
	line width = 0.95pt,
	style = solid,
	]
	table {final/out1_3.dat};
	\addlegendentry{Sim. Results}	

	\addplot[
	blue,
	no marks,
	line width = 0.95pt,
	style = solid,
	]
	table {final/out2_3.dat};
	\addplot[
	blue,
	no marks,
	line width = 0.95pt,
	style = solid,
	]
	table {final/out2_2.dat};
	\addplot[
	red,
	no marks,
	line width = 0.95pt,
	style = solid,
	]
	table {final/out3_3.dat};
	\end{semilogyaxis}
	\end{tikzpicture}
	\caption{OP versus transmit power.}
	\label{fig:outage_pt}
\end{figure}

Fig. \ref{fig:outage_pt} illustrates the OP versus the transmit power for a cascaded RIS with $=\{2,3\}$ and a single-RIS scenario. Specifically, we examine three different RIS-routes consisting of $\mathrm{RIS}_2$, $\mathrm{RIS}_1$-$\mathrm{RIS}_2$ and $\mathrm{RIS}_3$-$\mathrm{RIS}_1$-$\mathrm{RIS}_2$, which are denoted in Fig. 1 as Route 1, Route 2 and Route 3, respectively. For this case, the channel between the BS and $\mathrm{RIS}_2$ in Route 1, has shape parameter and spread parameter both equal to $1$, i.e., Rayleigh fading, and path loss exponent equal to $3.5$ and the BS-$\mathrm{RIS}_3$ channel is assumed to undergo Nakagami-$m$ fading with shape parameter equal to $3$ and spread parameter equal to $1$. It can be observed that the RIS-route with two RISs performs better in terms of OP compared to a single-RIS system, even if the phase estimation for the cascaded system is less accurate than the single-RIS scenario. However, the same does not hold if the participating RISs are three due to the multiplicative nature of the path loss term and the fact that the path loss term is greater than the achieved beamforming gain of the cascaded RIS system. Thus, the routing algorithms should consider the achieved beamforming gain and the resulting path loss for each RIS-route before selecting the participating RISs. 

\begin{figure}
	\centering
	\begin{tikzpicture}
	\begin{semilogyaxis}[
	width=0.82\linewidth,
	xlabel = {$P_t$ (dBm)},
	ylabel = {$C_c$ (bits/Hz)},
	xmin = 0,xmax = 60,
	ymin = 0.1,
	ymax = 50,
	xtick = {0,10,...,60},
	grid = major,
      legend style = {font = \tiny},
	legend cell align = {left},
	legend pos = south east
	]
	\addplot[
	black,
	only marks,
	mark=square*,
	mark repeat = 40,
	mark size = 1.5,
	]
	table {final/erg1_2.dat};
	\addlegendentry{$\kappa=2$, $L=1$}	
	\addplot[
	blue,
	only marks,
	mark=diamond*,
	mark repeat = 40,
	mark size = 2,
	]
	table {final/erg2_2.dat};
	\addlegendentry{$\kappa=2$, $L=2$}	
	\addplot[
	blue,
	only marks,
	mark=*,
	mark repeat = 40,
	mark size = 2,
	]
	table {final/erg2_3.dat};
	\addlegendentry{$\kappa=3$, $L=2$}	
	\addplot[
	red,
	only marks,
	mark=triangle*,
	mark repeat = 40,
	mark size = 2,
	]
	table {final/erg3_2.dat};
	\addlegendentry{$\kappa=2$, $L=3$}	
	\addplot[
	black,
	no marks,
	line width = 0.95pt,
	style = solid
	]
	table {final/erg1_2.dat};
	\addlegendentry{Sim. Results}	
	\addplot[
	blue,
	no marks,
	line width = 0.95pt,
	style = solid
	]
	table {final/erg2_2.dat};
	\addplot[
	blue,
	no marks,
	line width = 0.95pt,
	style = solid
	]
	table {final/erg2_3.dat};
	\addplot[
	red,
	no marks,
	line width = 0.95pt,
	style = solid
	]
	table {final/erg3_2.dat};
	\end{semilogyaxis}
	\end{tikzpicture}
	\caption{EC versus transmit power.}
	\label{fig:erg_pt}
\end{figure}
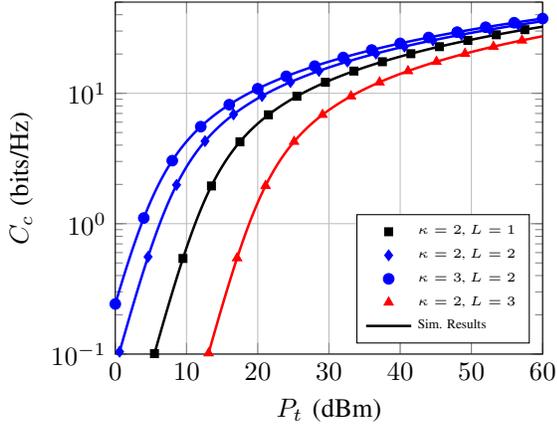

Fig. \ref{fig:erg_pt} depicts the performance of the considered network in terms of EC for each selected RIS-route. It can be observed that the achieved EC in a double-RIS system is greater than the EC of a single-RIS system, whereas the single-RIS system performs better than a cascaded RIS system with three RISs. Moreover, as the transmit power increases, the EC for all the presented scenarios converges at the same value, indicating that a cascaded RIS system can barely outperform a single-RIS system if the transmit power value is large. Therefore, by properly choosing the number of participating RIS in the selected RIS-route, the EC can be improved without increasing the transmit power.

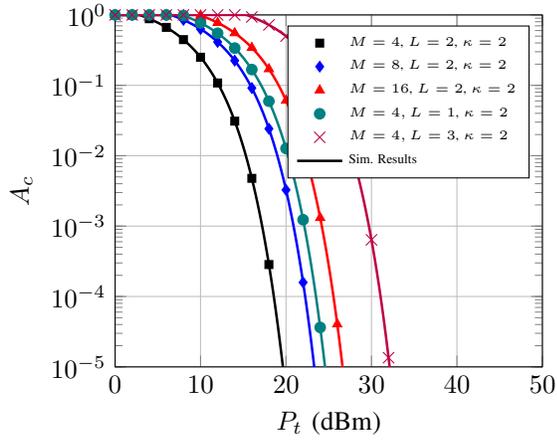
\begin{figure}
	\centering
	\begin{tikzpicture}
	\begin{semilogyaxis}[
	width=0.82\linewidth,
	xlabel = {$P_t$ (dBm)},
	ylabel = {$A_c$ },
	xmin = 0,xmax = 50,
	ymin = 0.00001,
	ymax = 1,
	xtick = {-20,-10,...,60},
	grid = major,
      legend style = {font = \tiny},
	legend cell align = {left},
	legend pos = north east
	]
	\addplot[
	black,
           only marks,
	mark=square*,
	mark repeat = 20,
	mark size = 1.5,
	]
	table {final/asep2_24.dat};
	\addlegendentry{$M=4$, $L=2$, $\kappa=2$}	
	\addplot[
	blue,
           only marks,
	mark=diamond*,
	mark repeat = 20,
	mark size = 2,
	]
	table {final/asep2_28.dat};
	\addlegendentry{$M=8$, $L=2$, $\kappa=2$}	
	\addplot[
	red,
           only marks,
	mark=triangle*,
	mark repeat = 20,
	mark size = 2,
	]
	table {final/asep2_216.dat};
	\addlegendentry{$M=16$, $L=2$, $\kappa=2$}	
	\addplot[
	teal,
           only marks,
	mark=*,
	mark repeat = 20,
	mark size = 2,
	]
	table {final/asep1_24.dat};
	\addlegendentry{$M=4,L=1$, $\kappa=2$}
          \addplot[
	purple,
          only marks,
           mark=x,
	mark repeat = 20,
	mark size = 3,
	]
	table {final/asep3_24.dat};
	\addlegendentry{$M=4,L=3$, $\kappa=2$}
	\addplot[
	black,
	no marks,
	line width = 0.95pt,
	style = solid,
	]
	table {final/asep2_24.dat};
	\addlegendentry{Sim. Results}
	\addplot[
	blue,
	no marks,
	line width = 0.95pt,
	style = solid,
	]
	table {final/asep2_28.dat};
	\addplot[
	red,
	no marks,
	line width = 0.95pt,
	style = solid,
	]
	table {final/asep2_216.dat};
	\addplot[
	teal,
	no marks,
	line width = 0.95pt,
	style = solid,
	]
	table {final/asep1_24.dat};
	\addplot[
	purple,
	no marks,
	line width = 0.95pt,
	style = solid,
	]
	table {final/asep3_24.dat};
	\end{semilogyaxis}
	\end{tikzpicture}
	\caption{ASEP versus transmit power for M-QAM. }
	\label{fig:asep_pt}
\end{figure}

Finally, in Fig. \ref{fig:asep_pt}, it is illustrated how the system's ASEP varies with the transmit power for the considered RIS-routes if the selected modulation scheme is QAM. As it can be observed, the utilization of two cascaded RISs can enhance the system's data transmission accuracy in an energy-efficient way. Specifically, it can be observed that the ASEP for $M=8$ in the double-RIS system is lower than the single-RIS system ASEP for $M=4$. Furthermore, as the modulation order increases, the ASEP deteriorates, highlighting the need for larger number of reflecting elements for reliable and energy-efficient communications with high-order constellations.
\vspace{-0.3cm}
\section{Conclusions} \label{conclusion}
In this work, we have characterized the performance of a cascaded RIS communication system by taking into account imperfect phase estimation. In more detail, we have examined the coverage capabilities of a single-user communication system aided by a cooperative multi-RIS route and its communication links average capacity and data transmission accuracy. Specifically, it has been shown that a double RIS system can improve the network's performance in terms of coverage and data transmission accuracy in an energy-efficient way, even if it is affected by imperfect phase estimation. Furthermore, it was shown that by increasing the number of participating RISs in the selected route for the presented communication system does not necessarily lead to performance improvement.

\vspace{-0.2cm}

\bibliographystyle{IEEEtran}
\bibliography{Bibliography}


\end{document}